\documentclass[11pt]{article}
\usepackage{graphicx}
\usepackage{epsfig}
\usepackage{epsf,amsfonts,amssymb}
\newcommand{\bea}{\begin{eqnarray}}
\newcommand{\eea}{\end{eqnarray}}

\begin{document}

\thispagestyle{empty}

{\bf \huge Euclidean Methods and the entropy function}\\

\vspace*{1cm}

\noindent{\bf Pedro J. Silva,}

\vspace*{0.5cm}

\noindent{\it Institut de Ci\`encies de l'Espai (IEEC-CSIC) and\\
Institut de F\'{\i}sica d'Altes Energies (IFAE),\\
E-08193 Bellaterra (Barcelona), Spain\\ psilva@ifae.es}\\[.3em]

\noindent{ \verb"Received 15 April 2008, Published in Fortschr. Phys. 56, No 7 - 9, 856 - 861 (2008)"}

\vspace{.5cm} {\bf ABSTRACT}

\vspace*{.5cm}

\noindent We review results of articles hep-th/0607056, hep-th/0610163 and 0704.1405 [hep-th].
Here we focus on establish the connection between the entropy
functional formalism of Sen and the standard Euclidean formalism taken at zero temperature. We find that Sen's entropy function $f$ (on-shell) matches the zero temperature limit
of the Euclidean action. Moreover, Sen's near horizon angular and
electric fields agree with the chemical potentials that are defined
from the zero-temperature limit of the Euclidean formalism. Connection
with the Dual CFT thermodynamics is briefly discussed.

\noindent

\section{Introduction}
This article contains the talk based on the articles \cite{SQSR1,SQSR2,Dias:2007dj}.
Here we only show the main results and general lessons that steam from our work.
More References and more details should be found on our original papers.

\section{On Sen`s entropy functional formalism}
Black holes (BH) are one of most interesting laboratories we have to
investigate quantum gravity effects. Due to their thermodynamic
behavior these objects have been associated to ensembles of
microstates in the fundamental quantum gravity theory where ideally,
quantum statistical analysis should account for all the BH
coarse-grained thermodynamical behavior. In particular, many
important insights in the classical and quantum structure of BH have
been obtained studying supersymmetric configurations in string
theory. In this context we have the so called attractor mechanism.
It was originally thought in the context of four
dimensional $N=2$ supergravity, where  we have that the values of
the scalar fields at the horizon are given by the values of the BH
conserved charges and are independent of the asymptotic values of
the scalars at infinity.

Importantly, the attractor mechanism has provided a new way to
calculate the BH entropy. In a series of articles
\cite{Sen:2005wa,Sen:2005iz,Astefanesei:2006dd}, Sen recovered the
entropy of  $D$-dimensional BPS BH using only the near horizon part
of the geometry. Basically, in this regime the solution adopts the
form $AdS_2 \otimes S^{D-2}$ \footnote{The analysis of the near
horizon geometry has been applied to more general BH that define
squashed $AdS_2\otimes S^{D-2}$ geometries.} plus some electric and magnetics
fields. The entropy $S$ is obtained by introducing a function $f$ as
the integral of the corresponding supergravity Lagrangian over the
$S^{D-2}$. More concretely, an entropy function is defined as $2\pi$
times the Legendre transform of $f$ with respect to the electric
fields $e_i$. Then, an extremization procedure fixes the on-shell
BPS values of the different fields of the solution and in particular
determines the BPS value of the entropy $S$,

Sen's entropy functional formalism assumes that: (i) we start with a
Lagrangian $\mathcal{L}$ with gravity plus some field strengths and
uncharge massless scalar fields; and (ii) due to the attractor
mechanism the near horizon geometry of a $D$-dimensional BH is set
to be of the form $AdS_2\otimes S^{D-2}$. From the above input data,
the general form of the near horizon BH solution is
\bea &&ds^2=v_1\left(-\rho^2d\tau+{d\rho^2\over \rho^2}\right)+v_2d\Omega_{D-2}^2\,,\nonumber\\
&&F^{(i)}_{\rho \tau}=e_i\,,\quad\quad\quad
H^{(a)}=p_a\epsilon_{D-2}\,,\nonumber\\
&&\phi_s=u_s\,, \label{geralNHsen} \eea
where $\epsilon_{D-2}$ is the unit-volume form of $S^{D-2}$, and
$(e_i,p_a)$ are respectively the electric fields and the magnetic
charges of the BH. Note that $(\vec u,\vec v, \vec e, \vec p)$ are
arbitrary constants up to now and therefore the solution is
off-shell. Next, it is defined the following function
\bea f(\vec u,\vec v,\vec e,\vec p)=\int_{S^{D-2}}\sqrt{-
g}\mathcal{L}\,, \label{geralfsen}  \eea
where $\mathcal{L}$ is the string frame Lagrangian of the theory.
After minimizing $f(\vec u,\vec
v,\vec e,\vec p)$ with respect to $(\vec u,\vec v)$ we obtain the
exact supersymmetric near horizon BH solution in terms of $(\vec e,
\vec p)$. In fact, the field equations are reproduced by this
minimization procedure. Furhermore, minimization with respect to
$\vec e$ gives the electric charges $\vec q$. Explicitly,  the
on-shell values of $\vec u,\vec v,\vec e$ that specify
(\ref{geralNHsen}) for a given theory described by (\ref{geralfsen})
are found through the relations,
\bea {\partial f\over \partial u_s}=0\,,\quad\quad {\partial f\over
\partial v_j}=0\,, \quad\quad {\partial f\over \partial e_i}=q_i\,.
\eea
Then, using Wald formalism \cite{Wald}, Sen derived that the entropy
$S$ of the corresponding BH is given by $2\pi$ times the Legendre
transform of $f$,
\bea S=2\pi\left(e_i{\partial f\over \partial e_i}-f \right)\,.
\label{sen-entropy}\eea
Finally notice that the minimization procedure, can be taken only
after $S$ is defined. In this form $S$ is really an entropy function
of $(\vec u,\vec v,\vec q,\vec p)$, that after minimization equals
the BH entropy as a function of $(\vec q,\vec p)$ only.

The above formalism fixes the form of the NH geometry and the entropy $S$
in terms of the conserved charges but what is the geometric origin or motivation for
the above definitions? and how is connected to the usual Bh thermodynamics?
To answer these questions we revisit the Bh thermodynamics and the limit
of zero temperature in next section.

\section{On GR thermodynamics and zero temperature limit}

In \cite{SQSR1,SQSR2} the ``thermodynamics" or better ``the
statistical mechanics" of supersymmetric solitons in gauged
supergravity was studied in detail using an extension of standard
Euclidean thermodynamical methods to zero temperature systems. We
call this approach the Euclidean zero-temperature formalism. BPS BH
can be studied as dual configurations of supersymmetric ensembles at
zero temperature {\it but} non-zero chemical potentials in the dual
CFT. These potentials control the expectation value of the
conjugated conserved charges carried by the BH, like {\it e.g.,} angular
momenta and electric charge.

In these articles, the two main ideas are: \verb"First", there is a
supersymmetric field theory dual to the supergravity theory.
\verb"Second", in this dual field theory the grand canonical
partition function over a given supersymmetric sector can be
obtained as the zero temperature limit of the general grand
canonical partition function at finite temperature. This limit also
fixes the values of several chemical potentials of the system.

To make things more clear, recall that all supersymmetric states in
a field theory saturate a BPS inequality that translates into a
series of constraints between the different physical charges. For
definiteness, let us consider a simple case where the BPS bound
corresponds to the constraint: $E=J$.
Then, defining the left and right variables $E^\pm=
\hbox{$1\over2$}(E_\nu\pm J_\nu)$, $\beta_\pm=\beta(1 \pm \Omega)$
the grand canonical partition function is given by
\bea Z_{(\beta,\Omega)}=\sum_\nu e^{-(\beta_+E_+ + \beta_-E_-)}\,.
\eea
At this point, it is clear that taking the limit $\beta_-\rightarrow
\infty$ while $\beta_+\rightarrow \omega$ (constant), gives the
correct supersymmetric partition function. The above limiting
procedure takes $T$ to zero, {\it but} also scales $\Omega$ in such
a way that the new supersymmetric conjugated variable $\omega$ is
finite and arbitrary. Note that among all available states, only
those that satisfy the BPS bound are not suppress in the sum,
resulting in the supersymmetric partition function
\bea Z{(\omega)}=\sum_{bps} e^{-\omega J}\,, \eea
where the sum is over all supersymmetric states ($bps$) with $E=J$.
The above manipulations are easy to implement in more complicated
supersymmetric field theories like, {\it e.g.}, $N=4$ SYM theory in
four dimensions. What is less trivial is that amazingly it could
also be implemented in the dual supersymmetric configurations of
gauged supergravity, since it means that these extreme BPS solutions
are somehow protected from higher string theory corrections.

To apply the Euclidean zero-temperature formalism to concrete
black hole systems, it is profitable to highlight its key steps. To
study the statistical mechanics of supersymmetric black holes we
take the off-BPS BH solution and we send $T\rightarrow 0$. In this
limiting procedure, the angular velocities and electric potentials
at the horizon can be written as an expansion in powers of the
temperature. More concretely one has when $T\rightarrow 0$,
\bea  \beta \rightarrow \infty\,,\quad\Omega \rightarrow
\Omega_{bps}\,-\,{\omega\over\beta}\,+ \,O(\beta^{-2})\,,\quad \Phi\rightarrow \Phi_{bps}\,-\,{\phi\over\beta}\,+ \,O(\beta^{-2})\,,
\label{multi}\eea
where $\beta$ is the inverse temperature; $(\Omega, \Phi)$ are the
angular velocities and electric potentials at the horizon; the
subscript $bps$ stands for the values of these quantities in the
on-shell BPS solution; and $(\omega,\phi)$ are what we call the
supersymmetric conjugated potentials, {\it i.e.}, the next to
leading order terms in the expansion. For all the systems studied,
we find that the charges have the off-BPS expansion,
\bea \label{gen1}  E=E^{bps}+\mathcal{O}\left( \beta^{-2} \right)
\,, \qquad Q = Q^{bps}+\mathcal{O}\left( \beta^{-2} \right)
\,,\qquad J = J_{\phi}^{bps} +\mathcal{O}\left( \beta^{-2}\right),
\eea
where $(E,Q,J)$ are the energy, charges and angular momenta of the
BH. In supergravity, the grand canonical partition function in the
saddle point approximation is related to so called quantum
statistical relation (QSR)
\bea I_{(\beta,\Phi,\Omega)}=\beta E-\Phi Q-\Omega J-S\,,
\label{QSR:intro}\eea
where $S$ is the entropy, and $(\beta,\Phi,\Omega)$ are interpreted
as conjugated potentials to $E,Q,J$, respectively. $I$ is the
Euclidean action (evaluated on the off-BPS BH solution) that, in
this ensemble, depends only on $(\beta,\Phi,\Omega)$. It plays the
role of free energy divided by the temperature. Inserting
(\ref{multi}) and (\ref{gen1}) into (\ref{QSR:intro}) yields
\bea I_{(\beta,\Phi,\Omega)}=\beta (E^{bps}-\Phi_{bps}
Q^{bps}-\Omega_{bps} J^{bps})+\phi Q^{bps}+\omega J^{bps}-S_{bps}  +
\mathcal{O}\left( \beta^{-1}\right)\,. \label{QSR2:intro}\eea
Here, we observe that this action is still being evaluated {\it
off}-BPS. Moreover, the term multiplying $\beta$ boils down to the
BPS relation between the charges of the system and thus vanishes
(this will become explicitly clear in the several examples we will
consider). This is an important feature, since now we can finally
take the  $\beta\rightarrow \infty$ limit yielding relation
\begin{equation}
I_{bps}=\phi Q^{bps}+\omega J^{bps}-S_{bps}\,.
\label{sqsr:intro}\end{equation}
It is important to stress that this zero temperature limiting
procedure yields a finite, not diverging, supersymmetric version of
QSR, or shortly SQSR. Note that if we had evaluated the Euclidean
action (\ref{QSR:intro}) directly on-shell it would not be well
defined, as is well-known. As a concrete realization, we picked (and
will do so along the paper) the SQSR to exemplify that the
$T\rightarrow 0$ limit yields well-behaved supersymmetric relations.
The reason being that this SQSR relation is the one that will
provide direct contact with Sen's entropy functional formalism,
which is the main aim of our study. However, it also provides a
suitable framework that extends to the study of the {\it full}
statistical mechanics of supersymmetric black holes.

\section{On entropy functional and zero temperature thermodynamics}

In previous sections we have described two apparently unrelated
procedures to obtain the entropy of supersymmetric BH that naturally
contain the definitions of pairs of conjugated variables, related to
the BH charges. In this section we show that both procedures produce
basically the same body of final definitions, even though
conceptually both approaches are rather different.

That both approaches produce the same final chemical potentials and
definitions can be seen in any of the examples at hand. As usual,
the best way to illustrate our point is to pick a system that
captures the fundamental ingredients.
Comparing the zero temperature thermodynamic relations
with the corresponding Sen's definitions, in any BH like
{\it e.g.,} the D1-D5-P system, we can indeed
confirm that all the key quantities agree in the two formalisms.
Explicitly we have that
\bea \phi_i=2\pi e_i\,,\quad\quad\quad Q_i=q_i\,,\quad\quad\quad
I_{bps}=2\pi f\,. \label{phi:e} \eea

Nevertheless, that both frameworks are equivalent is {\it a priori}
not at all obvious since they have important differences. Sen's
approach relies completely on the structure of the near horizon
geometry. In particular, the entropy is constructed analyzing Wald's
prescription and Einstein equations in these spacetimes and all the
analysis is carried on at the BPS bound {\it i.e.}, when the
solution is extremal. In contrast, the zero temperature limit
approach relies on the thermodynamical properties of BH and, in
principle, uses the whole spacetime, not only the near horizon
region. The resulting thermodynamic definitions come as a limiting
behavior of non-extremal BH and have a nice straightforward
interpretation in terms of the dual CFT thermodynamics.

\subsection{Near-horizon and
asymptotic contributions to the Euclidean action} \label{sec:Map1}

To understand why the above close relations between the two
formalisms hold, let us go back to the calculation of the Euclidean
action for general BH in the {\it off}-BPS regime. Inspired in ten
dimensional type II supergravity, we start with the general
action\footnote{For simplicity, the reasoning is done at the level
of two derivative Lagrangian. Nevertheless, following Wald's
approach for higher derivative actions, we notice that the BH action
can always be recast as surface integrals. Moreover, for
definiteness, we anchor our discussion to type II action, but
whenever needed we make comments to extend our arguments to more
general theories.}
\bea I={1\over 16\pi G}
\int_\Sigma{\sqrt{-g}\left(R-{1\over2}(\partial\Psi)^2-{1\over 2
n!}e^{\alpha\Psi}F_{(n)}^2\right)}+{1\over 8\pi
G}\int_{\partial\Sigma}{K}\,,\eea
where $\Sigma$ is the spacetime manifold, $\partial\Sigma$ the
boundary of that manifold and $K$ is the extrinsic curvature. In the
BH case, once we have switched to Euclidean regime, it is necessary
to compactify the time direction to avoid a conical singularity.
This compactification defines the Hawking temperature as the inverse
of the corresponding compactification radius.

To evaluate the Euclidean action on the BH solution, one of the
methods to obtain a finite result, {\it i.e.}, to regularize and
renormalize the action, consists of putting the BH in a box and
subtract the action of a background vacuum solution
$(g^0,\Psi^0,F^0)$. This procedures also defines the ``zero" of all
the conserved charges. For asymptotic flat solutions we use
Minkowski, while for asymptotic AdS solutions we use AdS. Once in
the box, the radial coordinate is restricted to the interval
$(r_+,r_b)$, where $r_+$ is the position of the horizon and $r_b$
corresponds to an arbitrary point which limits the box and that at
the end is sent to infinity. Another important ingredient is the
boundary conditions on the box. Basically, depending on which
conditions we impose on the different fields, we will have fixed
charges or fixed potentials. If we do not add any boundary term to
the above action, we will be working with fixed potentials, {\it
i.e.}, we will work in the grand canonical ensemble.

At this point we are ready to rewrite the Euclidean action in two
pieces, one evaluated in the first boundary at $r=r_+$, and the
other in the second boundary at $r=\infty$,
\bea I=\int_{r=r_+}\left\{ {c\over 8\pi G}e^{a\Psi}F_{(n)}C_{(n-1)}
+ {1\over 8\pi G}K\right\} + \int_{r=\infty}\left\{ {c\over 8\pi
G}e^{a\Psi}F_{(n)}C_{(n-1)} + {1\over 8\pi
G}\left(K-K^0\right)\right\} \,.  \eea
where the field equations have been used and $c$ is a proportionality constant.
Then we can rewrite the above expression as,
\bea I=\underbrace{\beta(\Phi_{bps}-\Phi)Q-S} \,+\,
\underbrace{\beta(E-\Phi_{bps}Q)}\,.\\
r=r_+\quad\quad\quad\quad\quad\: r=\infty\quad\quad \nonumber
\label{actionsplit}\eea
Therefore we can always find a gauge in which the Euclidean action
splits in two contributions, one at the horizon and the other in the
asymptotic region. In fact, from our discussion it is easy to see
that the first term exactly reproduces the SQSR, {\it i.e.},
\bea \lim_{BPS\;limit}\; \beta(\Phi_{bps}-\Phi)Q-S = \phi
Q_{bps}-S_{bps}\,. \label{split1}\eea
while the asymptotic term vanishes due to fact that
$\Phi_{bps}=1$, and thus the leading term in the expansion is
nothing else than the BPS relation $E_{bps}=Q_{bps}$ characteristic
of supersymmetric regimes, {\it i.e.},
\bea \lim_{BPS\;limit}\; \beta(E-\Phi_{bps}Q)=\lim_{BPS\;limit}\;
\beta(E-Q) = 0 \label{split2}\,.\eea
The above results are trivially generalized to the case of rotating
charged BH, see \cite{Dias:2007dj}, for more detail.

Hence, due to the above equations we have verified that
\bea I_{bps}=2\pi f\,.\eea
\subsection{Relation between chemical potentials in the two
formalisms} \label{sec:Map2}

At this point only reminds to understand the relation between the
conjugated potentials in both pictures. In Sen's approach, the
information about them is contained in the electric fields of the
near horizon geometry, while in the Euclidean zero temperature
formalism this information is encoded in the next to leading order
term in an off-BPS expansion of the full geometry. Although these
definitions seem to be rather different at first sight, notice that
in Sen's approach the field strength is just the radial derivative
of the potential evaluated at the horizon. In the Euclidean zero
temperature case, the off-BPS expansion can be rewritten as an
expansion in the radial position of the horizon $\rho_+$. Therefore,
the next to leading order term in the off-BPS expansion of the gauge
potential at $\rho_+$ is proportional to its derivative with respect
to the radial position of the horizon. Hence it is proportional to
the field strength at the horizon. It is not dificult to check that
the above reasoning produces exactly the result
\bea \phi_i=2\pi e_i \,.
\label{e:eq4} \eea

therefore we have been able to relate all the different terms in the
entropy functional approach with the emergent chemical potentials and thermodynamic
functions of the zero temperature Euclidean approach.

\section{Conclusions}

As stated above, the main goal of this article is to provide a
bridge between Sen's entropy functional formalism and standard
Euclidean analysis of the thermodynamics of a black hole system.
While doing so, we also find that the supergravity conjugated
potentials defined in Sen's formalism map into chemical potentials
of the dual CFT.

We obtain a unifying picture where:

\verb 1) We are able to recover the entropy function of Sen from the
zero temperature limit of the usual BH thermodynamics and the
statistical mechanics definitions of the dual CFT theory. The
supergravity and their dual CFT chemical potentials are identified
with the surviving Sen's near horizon electric and angular fields.
The Euclidean action is identified with Sen's function $2\pi f$.

\verb 2) As a byproduct of the above analysis we have understood how
to calculate the BPS chemical potentials that control the
statistical properties of the BH using only the BPS regime, {\it
i.e.}, without needing the knowledge of the non-BPS geometry. The
CFT chemical potentials are dual to the supergravity ones.
Traditionally, to compute the latter we have to start with the
non-BPS solution and send the temperature to zero to find the next
to leading order terms in the horizon angular velocities and
electric potentials expansions that give the chemical potentials.
This requires the knowledge of the non-BPS geometry. Unfortunately,
sometimes this is not available and we only know the BPS solution.
But, from item 1) we know that the near horizon fields, that Sen
computes with the single knowledge of the BPS near horizon solution,
give us the supergravity chemical potentials. So now we can compute
the supergravity chemical potentials of any BPS BH solution,
regardless of its embedding into a family of non-BPS solutions,
while still keeping the relation with the dual CFT.

\section*{Acknowledgments}

\noindent {\small The author would like to thanks the organizers of the 3rd RTN Workshop for
hospitality. This work was partially funded by the Ministerio de
Educacion y Ciencia under grant FPA2005-02211 and CSIC via the I3P programme}

\end{document}